# Exploring NoC Mapping Strategies: An Energy and Timing Aware Technique

César Marcon[1], Ney Calazans[2], Fernando Moraes[2], Altamiro Susin[1], Igor Reis[2], Fabiano Hessel[2]
[1] *PPGC - II - UFRGS - Av. Bento Gonçalves, 9500, Porto Alegre, RS – Brazil*
[2] *PPGCC - FACIN – PUCRS - Av. Ipiranga, 6681, Porto Alegre, RS – Brazil*
{marcon, susin}@inf.ufrgs.br,{calazans, moraes, ireis, hessel}@inf.pucrs.br

**Abstract**

*Complex applications implemented as Systems on Chip (SoCs) demand extensive use of system level modeling and validation. Their implementation gathers a large number of complex IP cores and advanced interconnection schemes, such as hierarchical bus architectures or networks on chip (NoCs). Modeling applications involves capturing its computation and communication characteristics. Previously proposed communication weighted models (CWM) consider only the application communication aspects. This work proposes a communication dependence and computation model (CDCM) that can simultaneously consider both aspects of an application. It presents a solution to the problem of mapping applications on regular NoCs while considering execution time and energy consumption. The use of CDCM is shown to provide estimated average reductions of 40% in execution time, and 20% in energy consumption, for current technologies.*

## 1. Introduction

SoC design requires special communication resources to fulfill stringent design requirements. Deep sub-micron effects pose formidable physical design challenges for long wires and global on-chip communication. Many designers have proposed a change from the mainstream synchronous design paradigm to a *globally asynchronous, locally synchronous* (GALS) paradigm [1]. In a GALS design, the application is partitioned into synchronous domains. Each domain is locally synchronous and placed inside a limited region, usually called *tile*. An asynchronous communication resource provides the link between synchronous domains. A NoC is an intra-chip communication infrastructure, usually composed by a set of routers interconnected by point to point communication channels, implementing a chosen topology. The NoC channels can be designed to provide an asynchronous communication protocol between otherwise synchronous domains. NoCs can then be easily adapted to implement systems based on the GALS paradigm. Besides, they present high scalability, reusability, and reliability [2].

Consider a SoC implemented using the GALS paradigm and composed by **n** cores. Suppose this SoC employs a NoC as its internal sole communication resource. The application mapping problem for this architecture consists in finding an association of each core to a tile (a *mapping*) such that some cost function is minimized.

In the most general case, the problem allows ***n!*** possible solutions. Given a future SoC with hundreds of tiles [3], exhaustive search of the problem solutions space will rapidly become unfeasible. Consequently, the optimal implementation of such SoCs requires efficient mapping strategies. Some mapping strategies have been proposed. For example, [4] and [5] propose a *communication weighted model* (CWM) to account for the overall communication volume of each channel. However, CWM does not consider communication timing. This paper proposes a *communication dependence and computation model* (CDCM) to capture both, the volume and timing of application communication. Comparing these models for a 0.07µ technology, CDCM produced estimated average reductions of 40% and 20% in execution time and energy consumption w.r.t. CWM.

The remaining of the paper is organized as follows. Section 2 discusses related work. Section 3 gives a formulation of the mapping problem. Section 4 describes and compares CWM and CDCM algorithms and Section 5 presents experimental results. Section 6 presents some conclusions and directions for further work.

## 2. Related Work

Hu and Marculescu [4] showed that by using mapping algorithms it is possible to reduce by more than 60% the energy consumption when compared to random mapping solutions. The authors propose the use of an *application characterization graph* (APCG), a way of capturing CWMs. Murali and De Micheli [5], propose a solution similar to that in [4]. Their CWM is represented by a structure called *core graph*. The main goal of their work is to propose an algorithm that maps cores on mesh NoC architectures under bandwidth constraints, trying to minimize average communication delay.

Ye et al. [6], propose a model to evaluate the energy consumption in a communication infrastructure containing routers, internal buffers and interconnected wires. The same authors [7] describe the contention problem in NoCs, evaluating the associated performance reduction. They propose as solution a routing algorithm that minimizes energy consumption, by reducing the required buffers in the communication network.

The approach of the present paper proposes and uses CDCM, which models application packet dependence and computation time. It explores CDCM and CWM strategies to solve mapping problems fulfilling energy consumption and execution time requirements. It stresses that CWMs as the ones presented in [4][5] abstract communication timing, an essential information to estimate execution time and energy consumption of the application. CDCMs lead to a better mapping solution if compared to CWMs, with low extra computational effort. Moreover, the dynamic energy model presented in [6] is extended here to include static energy consumption. Instead of evaluating different routing algorithms, as in [7], this work explores communication dependence graphs to reduce communication buffers, saving area, execution time and energy.

## 3. Problem Formulation

Each core behavior can be modeled by its computation and communication characteristics. Here a formulation of the mapping problem is proposed using graph structures, together with models, strategies, and algorithms for it.

### 3.1 Graph Definitions

Graph structures used in this work are defined here: CWG, CDCG, and CRG. Figure 1 illustrates them.

**Definition 1**: A *communication weighted graph* (CWG) is a directed graph $<C, W>$. The set of vertices $C = \{c_1, c_2…, c_n\}$ represents the set of cores in one application. Assuming $w_{ab}$ is the number of bits of all packets sent from a core $c_a$ to a core $c_b$, then the set of edges $W$ is $\{(c_a, c_b) \mid c_a, c_b \in C$ and $w_{ab} \neq 0\}$, and each edge is labeled with the value $w_{ab}$.

$W$ represents all communications between application cores, while CWG reveals information of application relative communication volume. This is similar to the definitions of *APCG* from [4] and *core graph* from [5].

**Definition 2**: Let $C$ be a set of cores of a given application. The *communication dependence and computation graph* or CDCG of this application is a directed graph $<P, D>$. The set of vertices P contains all packets exchanged between every pair of cores communicating in an application. There are also two special vertices named *Start* and *End*. The set of edges D contains all communications dependences in an application. Elements of P are 4-tuples in the form $p_{abq} = (c_a, c_b, t_{aq}, w_{abq})$, where $c_a, c_b \in C$, and $p_{abq}$ is the q-th packet sent from $c_a$ to $c_b$. This packet contains $w_{abq}$ bits and is transmitted after the computation time $t_{aq}$ of the originating core ($c_a$) has elapsed.

The set of all packets sent from $c_a$ to $c_b$ is $P_{ab}$.

The CDCG represents the communication and computation for an application composed by an arbitrary number of cores. The direction of the edges in this graph denotes that the destination vertex computation depends on the computation of the origin vertex. In other words, the destination vertex presents a *communication dependence* w.r.t. the origin vertex.

CWM and CDCM are evaluated here using a mesh topology NoC using wormhole, deterministic XY routing algorithm. Other NoC topologies can be equally treated.

**Definition 3**: A *communication resource graph* is a directed graph $CRG = <\Gamma, L>$, where the vertex set is the set of tiles $\Gamma = \{\tau_1, \tau_2, …, \tau_n\}$, and the edge set $L = \{(\tau_i, \tau_j), \forall \tau_i, \tau_j \in \Gamma\}$ gives the set of paths from $\tau_i$ to $\tau_j$.

The value $n$ is again the total number of tiles and is equal to $\varphi \times \omega$, the product of the two NoC dimensions. CRG edges and vertices represent physical links and routers of the target architecture, respectively. The CRG definition is equivalent to the *architecture characterization graph* in [4] and to the *NoC topology graph* in [5].

Figure 1 illustrates the above definitions using a hypothetical application with four IP cores exchanging a total of six packets and a 2×2 NoC. Figure 1(a) shows a CWG where $C = \{A,B,E,F\}$, the edge labels are $w_{AB} = 15$, $w_{AF} = 15$, $w_{BF} = 40$, $w_{EA} = 35$, $w_{FB} = 15$ and the set W can be extracted easily from the Figure. Figure 1(b) depicts one possible CDCG for the application, where $P = \{p_{EA1} = (E,A,10,20), p_{EA2} = (E,A,20,15), p_{AF1} = (A,F,6,15),…\}$ and $D = \{(Start,p_{EA1}), (p_{EA1}, p_{EA2}), (p_{AB1}, p_{AF1})…\}$. Figure 1(c) and Figure 1(d) depict two arbitrary mappings of $C$, each corresponding to a CRG as follows: (c) $CRG_1 = <\{\tau_1, \tau_2, \tau_3, \tau_4\}, \{(\tau_1,B),(\tau_2,A),(\tau_3,F),(\tau_4,E)\}>$ and (d) $CRG_2 = <\{\tau_1, \tau_2, \tau_3, \tau_4\},\{(\tau_1,B),(\tau_2,E),(\tau_3,F), (\tau_4,A)\}>$.

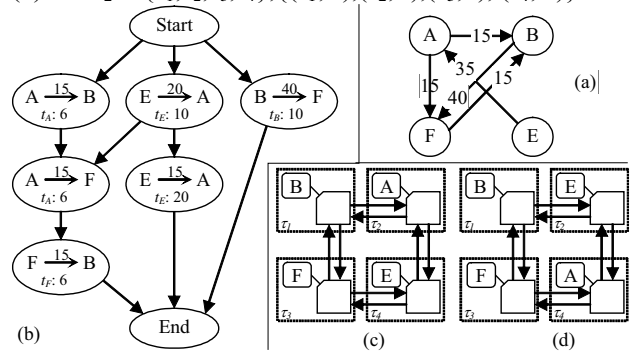

**Figure 1 – (a) CWG; (b) CDCG; (c,d) 2 mappings.**

### 3.2 Energy and Timing Model

Energy consumption originates from both IP cores and interconnection. This work focuses on NoC energy consumption, proposing models to estimate energy dissipation, used as cost functions to evaluate mappings.

Dynamic energy consumption is proportional to switching activity, arising from packets moving across the NoC, dissipating energy on the interconnect wires and inside each router. Static energy consumption comes mainly from leakage current, and is proportional to the application execution time and to the number of gates. Normally, static energy contributes with the smallest part of the energy consumption. However, for deep-submicron technologies, leakage current cannot be neglected, and static energy consumption is a significant part of the total energy consumption, reaching up to 20% in new technologies [8].

While CWMs are only suitable to compute dynamic energy, CDCMs can be used to estimate the total energy consumption of the NoC, enabling the computation of static and dynamic power dissipation figures. The approach here is similar to those in [4] and [5], but some concepts are extended to consider static energy consumption. The concept of bit energy $EBit$ [6] is used to estimate the dynamic energy consumption for each bit, when these flip its polarity from a previous value. $EBit$ is split into three components: bit dynamic energy ($ERbit$) dissipated at the router (router wires, buffers and logic gates); bit dynamic energy dissipated on horizontal ($ELHbit$) and vertical ($ELVbit$) links between tiles; and bit dynamic energy ($ECbit$) dissipated on the links between the router and the core of the tile. The relationship between these quantities is expressed by Equation (1), which computes the dynamic energy consumption of a bit passing through a router and a vertical or horizontal link.

(1) $EBit = ERbit + (ELHbit$ or $ELVbit) + ECbit$

$ERbit$ depends on the buffer structure and technology to estimate how many bit flips occur to write, read and preserve the information. $ELbit$ is directly proportional to the tile dimension. For regular mesh NoCs with square tiles, it is reasonable to estimate that $ELHbit$ and $ELVbit$ have the same value. Therefore, $ELHbit$ and $ELVbit$ are represented simply by $ELbit$. Furthermore, for large tile dimensions, $ECbit$ is negligible w.r.t. $ELbit$.

Equation (2) computes the dynamic energy consumed by a single bit traversing the NoC, from tile $\tau_i$ to tile $\tau_j$, where $\eta$ corresponds to the number of routers through which the bit passes.

(2) $EBit_{ij} = \eta \times ERbit + (\eta - 1) \times ELbit$

Let $\tau_i$ and $\tau_j$ be the tiles to which $c_a$ and $c_b$, are respectively mapped. Then, the dynamic energy consumed by a $c_a \rightarrow c_b$ communication is given by $EBit_{ab} = w_{ab} \times EBit_{ij}$. Equation (3) gives the total amount of *NoC dynamic energy consumption* ($EDyNoC$) for CWM, computing this for all bits of all communications ($y$) occurring in the NoC.

(3) $EDyNoC_{(CWM)} = \sum_{i=1}^{y} EBit_{ab}(i), \forall c_a, c_b \in C$

The *total NoC energy consumption* ($ENoC$) using CWM corresponds to $EDyNoC_{(CWM)}$, since CWM is a model that does not carry timing information. It should be clear from this reasoning that CWM is inappropriate to compute static energy consumption.

Let $w_{abq}$ be the total amount of bits of a packet $p_{abq} \in P_{ab}$, going from core $c_a$ to core $c_b$. Then, the dynamic energy consumed by the q-th packet of a $c_a \rightarrow c_b$ communication is given by $EBit_{abq} = w_{abq} \times EBit_{ij}$. Hence, equation (4) gives $EDyNoC$ for CDCM and $k_{abi}$ represents the number of packets of the i-th communication from core $c_a$ to core $c_b$.

(4) $EDyNoC_{(CDCM)} = \sum_{i=1}^{y} \sum_{q=1}^{k_{abi}} EBit_{abq}(i) \; \forall c_a, c_b \in C$

The *static power of each router* ($PSRouter$) is proportional to the number of gates that compose the router and that can be estimated by electrical simulation. With $n$ representing the number of tiles, equation (5) computes *NoC static power consumption* ($PstNoC$).

(5) $PStNoC = n \times PSRouter$

The *total packet delay* of the wormhole routing algorithm is composed by the *routing delay* and by the *packet delay*. The routing delay is the time necessary to create the communication path, which is determined during the traffic of the first flit, where the header of the packet is placed. The packet delay depends on the number of remaining flits. Let $n_{abq}$ be the number of flits of the q-th packet from $c_a$ to $c_b$, obtained by dividing $w_{abq}$ by the link width. Let $\lambda$ be the period of a clock cycle, and let $t_r$ be the number of cycles needed for taking a routing decision inside a router. Also, let $t_l$ be the number of cycles needed to transmit a flit through a link (between tiles or between an IP core and a router). The routing delay ($d_{Rijq}$) and the packet delay ($d_{Pijq}$) of the q-th packet from $\tau_i$ to $\tau_j$, are represented in Equations (6) and (7), considering that a packet goes through $\eta$ routers without contention. Contentions can only be determined at execution time.

(6) $d_{Rijq} = (\eta \times (t_r + t_l) + t_l) \times \lambda$

(7) $d_{Pijq} = (t_l \times (n_{abq} - 1)) \times \lambda$

The total packet delay ($d_{ijq}$), obtained from the sum of ($d_{Rijq}$) and ($d_{Pijq}$), is expressed by Equation (8).

(8) $d_{ijq} = (\eta (t_r + t_l) + t_l \times n_{abq}) \lambda$

The *application execution time* ($t_{exec}$) is obtained from both the computation of all $t_{iq}$ and $d_{ijq}$, and the contention time. Static energy consumption is proportional to $PstNoC$ and to $t_{exec}$. Thus, equation (9) computes *NoC static energy consumption* ($EstNoC$).

(9) $EStNoC = PStNoC \times texec$

Finally, equation (10) gives the overall static plus dynamic energy consumption at the NoC ($ENoC$) for CDCM.

(10) $E_{NoC(CDCM)} = E_{StNoC} + E_{DyNoC(CDCM)}$

## 4. Communication Algorithms Comparison

The FRW framework implements a simulated annealing search method to obtain mapping solutions for CWM and CDCM. Moreover, it can also execute an exhaustive search method to compare the quality of solutions against an absolute optimum solution, for small NoCs. Both algorithms, *CWM* and *CDCM*, start from an initial mapping, evaluate the mapping cost, and search for a new mapping that reduces the computed cost, until reaching a stop condition. For both algorithms, the mapping cost is stored inside cost variables of CRG edges and vertices. The sum of all cost variables determines the mapping objective function. Initially, all cores of *C* are randomly mapped onto the set of tiles $\Gamma$ and all cost variables are set to 0.

Given a mapping function, let tiles $\tau_i$ and $\tau_j$ be the respective images of cores $c_a$ and $c_b$ in this function. For the CWM algorithm, each total number of bits in all packets of $c_a \rightarrow c_b$ communications ($w_{ab}$) is associated to the cost variable of the corresponding vertices and edges of CRG. These start at $\tau_i$, follow the path defined by the XY routing algorithm and end at $\tau_j$. Equation (3) is the objective function of CWM, used to evaluate the cost of each mapping. The cost variable of each CRG edge is used to compute the dynamic energy of a link by multiplying $w_{ab}$ by $E_{Lbit}$. The cost variable of each CRG vertex is used to compute the dynamic energy of a router by multiplying $w_{ab}$ by $E_{Rbit}$. The sum of all cost variables of CRG results in $E_{DyNoC}$. This procedure is illustrated in Figure 2. CWM aims to find mappings reducing $E_{DyNoC}$.

CDCG improves CWM, as it considers computation time *and* packet ordering. Dependent packets cannot be concurrent. However, independent packets can occur at the same time, and may consequently lead to package contention, if they share the same communication resource. Packet contention implies larger *texec*, leading to increased $E_{StNoC}$. The CDCM algorithm searches for mappings that minimize the sharing of communication resources for concurrent packets. To allow this, each CRG edge and vertex are associated to a cost variable list, where each element of the list represents a packet that contains the number of bits, the absolute time interval that the packet is occupying the NoC resource, the source and target tiles.

The CDCM algorithm starts with all vertices pointed by the **Start** vertex (Figure 1(b)). Pointing to a CDCG vertex implies that the packet enclosed into the vertex may be *executed* onto the CRG. A *vertex execution* marks all of its output edges as free. A vertex can only be executed if all of its input edges are free. The algorithm searches for all free vertices following all paths, until all paths reach the **End** vertex. This procedure computes the time of each dependent path in the CRG. To execute a CDCG vertex that encloses the q-th packet onto the CRG, all bits of the q-th packet of $c_a \rightarrow c_b$ communication ($w_{abq}$) are associated with the corresponding vertices and edges of CRG, starting from $\tau_i$, following the XY routing algorithm, and ending in $\tau_j$.

When two or more packets compete for the same resource, all concurrent packets have to be contained into router input buffers. Therefore, from the contention point until reaching the target tile, contention time is added to the elapsed time, enabling to estimate the total packet delay more accurately. This procedure is illustrated in Figure 3 and Figure 4, where the communication weight of A→F is stored into the buffer of router $\tau_1$. The cost variable list of CRG edges and vertices computes the dynamic energy of links and routers by multiplying $w_{abq}$ by $E_{Lbit}$ and by $E_{Rbit}$, respectively. For a given mapping, the sum of all cost variables generates the value $E_{DyNoC}$. When all packets of CDCG are executed into CRG *texec* is obtained, enabling to compute $E_{NoC}$ through equation (10), the objective function to evaluate the mapping cost. The goal of CDCM is to find mappings that minimize $E_{NoC}$.

For both, CWM and CDCM algorithms, if the mapping cost achieved with a new mapping is smaller than the previously stored, the current mapping and cost are saved for further comparison. If the stop condition has not been reached, a new mapping is chosen and cost re-evaluated.

CWM and CDCM are able to estimate $E_{DyNoC}$ accurately, since this value depends only on the bit traffic along the NoC. The essential difference between the models is that CWM is not appropriate to estimate *texec*, due to the absence of task computation time and the impossibility of evaluating packet contention. The main advantages of CWM are (*i*) easy extraction of application core graph (CWG), since this can be done by simulation techniques; and (*ii*) low computational complexity. The automatic extraction of the CDCM application core graph is a hard task, since simulations only allow extracting possible message orderings, and not the required message dependence. As a consequence, CDCGs are described by hand, increasing the error susceptibility. The greater complexity of CDCM is observed on its algorithmic implementation, which increases computation time and memory usage w.r.t. CWM. However, CDCM captures the message ordering, and thus allows estimating the absolute execution time, and consequently the $E_{StNoC}$. The main drawback of CDCM evaluation is that in real embedded applications, the number of packets between cores is much larger than the number of cores. Since each vertex of CDCG represents a packet exchanged between two cores and each vertex of CWG represents a core, CDCGs are larger than CWGs. A comparison between the CPU time required by CDCM and CWM algorithms is presented in Section 5.

### 4.1 Energy and Timing Analysis

This Section shows the application of algorithms

CWM and CDCM to the example application mappings of Figure 1. The CWM evaluation does not capture the differences between the two mappings. However, CDCM evaluation shows that these mappings imply different execution times and distinct energy consumption estimates.

Figure 2 illustrates how the internal model of the CWM algorithm represents the energy for the two mappings of Figure 1(c, d). For illustration purposes, the example assumes that $ERbit = ELbit = 1 \times 10^{-12}$J/bit. The mapping of CWG onto CRG results that each vertex and edge of CRG is annotated with $ERbit$ and $ELbit$ energies, respectively. For instance, the execution of E→A communication of Figure 1(a) onto the mapping of Figure 2(a) implies $35 \times 10^{-12}$J of energy consumption, which is computed in tiles $\tau_4$ and $\tau_2$, and in the link between these tiles. Applying equation (3), over all communications of Figure 1(a), $EDyNoC$ results in $390 \times 10^{-12}$J for both mappings.

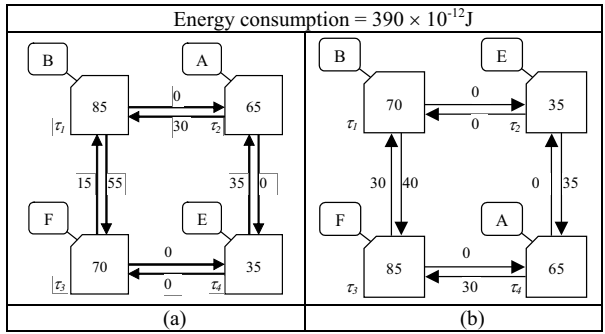

**Figure 2 – Mappings for Figure 1(c, d) - estimation of energy consumption obtained with CWM.**

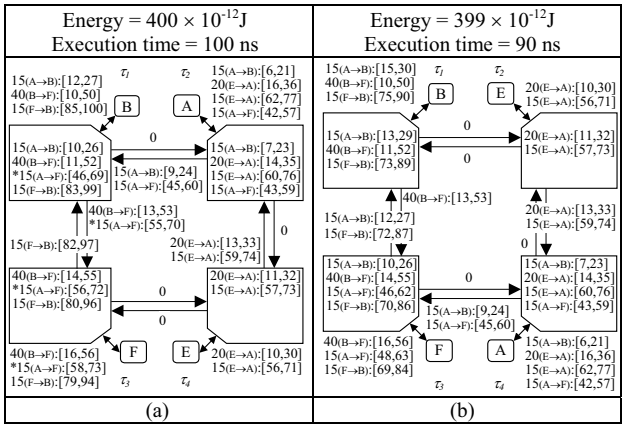

**Figure 3 – Figure 1(c, d) mappings, using CDCM.**

To evaluate CDCM mappings, consider the same parameters stated before, $t_r = 2$ clock cycles, $t_l = 1$ clock cycle, $\lambda = 1$ ns, one-bit sized flits, and unbounded router buffers. Figure 3 shows the same mappings illustrated before, now evaluated with CDCM. Each edge and each vertex is annotated with the number of bits in a given time interval. For instance, in Figure 3(a) tile $\tau_4$ is annotated with 20(E→A):[11,32] and 15(E→A):[57,73] which means that there are two packets from E to A. The first packet has 20 bits and occupies router $\tau_4$ during the interval from 11 ns to 32 ns. This interval is obtained by adding the E computation time $t_{E1} = 10$ ns with 1 ns of the link time. The link from $\tau_4$ to $\tau_2$ is annotated with the same two packets, each one delayed by the router delay.

Figure 4 depicts the CDCM algorithm execution over the mapping of Figure 3(a), showing a timing diagram that illustrates all computations and all packet deliveries. During A→F and B→F packets transmission some contentions occur, since they compete for the same resources at the same time. When the B→F packet uses the router of tile $\tau_1$, the A→F packet is contained into the input buffer of $\tau_1$. This contention implies that the packet is delayed until the B→F packet is transmitted to tile $\tau_3$. The effect of this delay is observed in the router variables of $\tau_1$ and the variables of the link between the $\tau_3$ router and IP core F. These variables are marked with '*' in Figure 3(a).

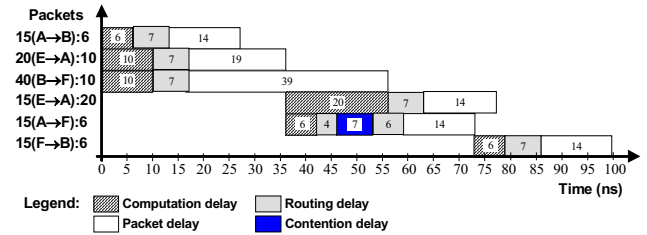

**Figure 4 – Timing for Figure 3(a) mapping.**

Figure 5 shows the timing diagram corresponding to the mapping depicted in Figure 3(b). It is possible to observe that this mapping avoids contention, since there are no packets competing for the same link at the same time. Changing the mapping of Figure 3(a) to that of Figure 3(b) implies an execution time reduction of 11.1%, from 100 ns to 90 ns. As there are different execution times for both mappings, the consumed static energy is different, too. For instance, consider $PstNoC=0.1 \times 10^{-12}$J/ns. Applying equation (10), $ENoC$ can be obtained for both mappings, showing that mapping (a) consumes 1 % more energy than (b). This difference cannot be computed by CWM, since it just captures the effects of dynamic energy.

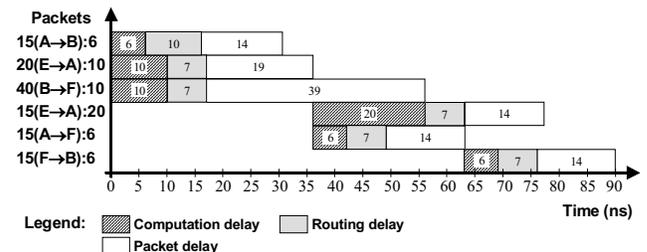

**Figure 5 – Timing for Figure 3(b) mapping.**

## 5. Experimental Results

Table 1 summarizes the characteristics of 18 applica-tions mapped onto 8 different NoC sizes. There are 4 embedded applications (a distributed Romberg integration, an 8-point Fast Fourier Transform, and 2 image applications for object recognition and image encoding) with some variations, for a total of 8 embedded applications. The remaining applications are benchmarks randomly generated by a proprietary system, which is similar to TGFF [9]; however, the system describes benchmarks through CDCGs, representing message dependence and bit volume of each message. The NoC size is the number of CRG vertices, the number of cores corresponds to the number of CWG vertices, and the number of packets of all cores equals the number of CDCG vertices.

**Table 1 – Summary of NoC/application features.**

| NoC size | Number of cores | Number of packets of all cores | Total volume of bits during application execution |
|---|---|---|---|
| 3 x 2 | 5; 6; 6 | 43; 17; 43 | 78,817; 174; 49,003 |
| 2 x 4 | 5; 7;8 | 16; 33; 18 | 1,600; 23,235; 5,930 |
| 3 x 3 | 7; 9; 9 | 16; 18; 32 | 1,600; 1,860; 43,120 |
| 2 x 5 | 8; 9; 10 | 24; 51; 22 | 2,215; 23,244; 322,221 |
| 3 x 4 | 10; 12; 14 | 15; 25; 88 | 3,100; 2,578,920; 115,778 |
| 8 x 8 | 62 | 344 | 9,799,200 |
| 10 x 10 | 93 | 415 | 562,565,990 |
| 12 x 10 | 99 | 446 | 680,006,120 |

For each application, the best mapping achieved with the CWM algorithm is compared to the best mapping achieved with the CDCM algorithm, and the results are summarized in Table 2. For both models exhaustive search (ES) and simulated annealing (SA) were applied, depending on the NoC size. ETR gives the *average execution time reduction*, and ECS denotes the *average energy consumption saving*, for a given technology, when mappings achieved with the CDCM algorithm are compared to the ones achieved with the CWM algorithm. $ECS_{0.35}$ column refers to ECS values obtained from 0.35μ technology, and $ECS_{0.07}$ column refers to ECS values obtained by estimating 0.07μ technology [8].

**Table 2 – Average energy and execution time reductions for CWM and CDCM.**

| Algorithm | NoC size | ETR | $ECS_{0.35}$ | $ECS_{0.07}$ |
|---|---|---|---|---|
| | 3 x 2 | 36 % | 0,50 % | 15 % |
| | 2 x 4 | 27 % | 0,43 % | 13 % |
| | 3 x 3 | 39 % | 0,55 % | 17 % |
| | 2 x 5 | 42 % | 0,72 % | 23 % |
| | 3 x 4 | 42 % | 0,71 % | 22 % |
| Simulated annealing only | 8 x 8 | 38 % | 0,6 % | 19 % |
| | 10 x 10 | 46 % | 0,8 % | 25 % |
| | 12 x 10 | 48 % | 0,86 % | 26 % |
| Average | | 40 % | 0,65 % | 20 % |

First, for small NoC sizes (up to 3x4 or 2x5), both ES and SA methods reached the same results. For larger NoC sizes (8x8, 10x10 and 12x10), it is not possible to find optimum mappings with ES within a reasonable time. The ETR column shows that the CDCM algorithm results in 40% average reduction of execution time when compared to the CWM algorithm. The $ECS_{0.35}$ column illustrates a very small energy consumption saving, since the leakage current is not as important for this technology. However, for deep-submicron technologies, there is a significant reduction in energy consumption (20% in average), as observable in column $ECS_{0.07}$. In addition, Table 2 shows a slight trend of energy consumption saving and execution time reduction when the NoC size increases.

The computational complexity of the CWM algorithm is proportional to the *number of communications between cores* (NCC) and the computational complexity of CDCM algorithm is proportional to the *number of dependences and packets of all cores* (NDP). In real embedded applications, NDP is larger than NCC. However, the increase in CPU time with the increase of the NDP/NCC ratio is approximately linear with a small slope. The worst case for CDCM took only 23% more CPU time than for CWM.

## 6. Conclusions and Future Work

This paper addressed the problem of mapping applications onto NoCs. A communication dependence and computation model (CDCM) has been introduced and compared to the communication weighted model (CWM). As a conclusion CDCM is able to reduce the application execution time and energy consumption when compared to CWM. Experimental results show an average of 40% in execution time reduction. CDCM also reduces energy consumption. For a 0.07μ technology, an average of 20% in energy savings is obtained. Moreover, CDCM presents only a moderate increase in computational cost, when compared to CWM, with better mapping results.

## References

[1] A. Iyer and D. Marculescu. *Power and performance evaluation of globally asynchronous locally synchronous processors*. **ISCA**, pp.158-168, May 2002.
[2] W. Dally and B. Towles. *Route packets, not wires: on-chip interconnection networks*. **DAC**, Jun. 2001.
[3] S. Kumar et al. *A network on chip architecture and design methodology*. **ISVLSI**, pp.105-112, Apr. 2002.
[4] J. Hu and R. Marculescu. *Energy-aware mapping for tile-based NoC architectures under performance constraints*. **ASP-DAC**, pp.233-239, Jan. 2003.
[5] S. Murali and G. De Micheli. *Bandwidth-constrained mapping of cores onto NoC architectures*. **DATE**, pp.896-901, Feb. 2004.
[6] T. T. Ye; L. Benini and G. De Micheli. *Analysis of power consumption on switch fabrics in network routers*. **DAC**, pp.524-529, Jun. 2002.
[7] T. T. Ye; L. Benini and G. De Micheli. *Packetization and routing analysis of on-chip multiprocessor networks*. **JSA**, vol. 50, issues 2-3, pp.81-104, Feb. 2004.
[8] D. Duarte et al. *Impact of scaling on the effectiveness of dynamic power reduction schemes*. **ICCD**, Sep. 2002.
[9] R. Dick, D. Rhodes and W. Wolf. *TGFF: task graphs for free*. **CODES/CASHE**, pp.97–101, Mar. 1998.